# Review: Semiconductor Quantum Light Sources

**Andrew J Shields**

Toshiba Research Europe Limited, 260 Cambridge Science Park, Cambridge CB4 0WE, UK

**Abstract**

Lasers and LEDs display a statistical distribution in the number of photons emitted in a given time interval. New applications exploiting the quantum properties of light require sources for which either individual photons, or pairs, are generated in a regulated stream. Here we review recent research on single-photon sources based on the emission of a single semiconductor quantum dot. In just a few years remarkable progress has been made in generating indistinguishable single-photons and entangled photon pairs using such structures. It suggests it may be possible to realise compact, robust, LED-like semiconductor devices for quantum light generation.

## Applications of Quantum Photonics

Applying quantum light states to photonic applications allows functionalities that are not possible using 'ordinary' classical light. For example, carrying information with single-photons provides a means to test the secrecy of optical communications, which could soon be applied to the problem of sharing digital cryptographic keys.[1,2] Although secure quantum key distribution systems based on weak laser pulses have already been realised for simple point-to-point links, true single-photon sources would improve their performance.[3] Furthermore, quantum light sources are important for future quantum communication protocols such as quantum teleportation.[4] Here quantum networks sharing entanglement could be used to distribute keys over longer distance or through more complex topologies.[5]

A natural progression would be to use photons for quantum information processing, as well as communication. In this regard it is relatively straightforward to encode and manipulate quantum information on a photon. On the other hand, single-photons do not interact strongly with one-another, a prerequisite for a simple photon logic gate. In linear optics quantum computing[6,7] (LOQC) this problem is solved using projective measurements to induce an effective interaction between the photons. Here triggered sources of single-photons and entangled pairs are required as both the qubit carriers, as well as auxiliary sources to test the successful operation of the gates. Although the component requirements for LOQC are challenging, they have recently been relaxed significantly by new theoretical schemes.[7] Quantum light states are also likely to become increasingly important for various types of precision optical measurement.[8]

For these applications we would ideally like light sources which generate pure single-photon states "on demand" in response to an external trigger signal. Key performance measures for such a source are the efficiency, defined as the fraction of photons collected into the experiment or application per trigger, and the second order correlation function at zero delay, see text box. The latter is essentially a measure of the two-photon rate compared to a classical source with random emission times of the same average intensity. In order to construct applications involving more than one photon, it is also important that photons emitted from the source (at different times), as well as those from different sources, are otherwise indistinguishable.

In the absence of a convenient triggered single-photon source, most experiments in quantum optics rely on non-linear optical processes for generating quantum light states. Optically pumping a crystal with a $\chi(2)$ non-linearity has a finite probability of generating a pair of lower energy photons via parametric down conversion. This may be used to prepare photon pairs with time-bin entanglement,[9] entangled polarisations,[10,11] or alternatively single-photon states 'heralded' by the second photon in the pair.[12] A $\chi(3)$ non-linearity in a semiconductor has also been used to generate entangled pairs.[13] As these non-linear processes occur randomly, there is always a finite probability of generating two pairs that increases with pump power. As double pairs degrade the fidelity of quantum optical gates, the pump laser power must be restricted to reduce the rate of double pairs to an acceptable level, which has a detrimental effect upon the efficiency of the source.[14] This means that although down-conversion sources continue to be highly successful in demonstrating few photon quantum optical gates, scaling to large numbers may be problematic. Solutions have been proposed based on switching multiple sources,[15] or storing photons in a switched fibre loop.[16]

Ideally we would like a quantum light source that generates exactly one single-photon, or entangled-pair, per excitation trigger pulse. This may be achieved using the emission of a single quantum system. After relaxation, a quantum system is by definition no longer excited and therefore unable to re-emit. Photon anti-bunching, the tendency of a quantum source to emit photons separated in time, was first demonstrated in the resonance fluorescence of a low density vapour of Na atoms,[17] and subsequently for a single ion.[18]

Quantum dots are often referred to as "artificial atoms", as their electron motion is quantised in all three spatial directions, resulting in a discrete energy level spectrum, like that of an atom. They provide a quantum system which can be grown within robust, monolithic semiconductor devices and can be engineered to have a wide range of desired properties. In the following we review recent progress towards the realisation of a semiconductor technology for quantum photonics. An excellent account of the early work can be found in Ref.[19]. Space restrictions limit discussion of work on other quantised systems. For this we refer the reader to the comprehensive review in Ref[20].

## Optical Properties of Single Quantum Dots

Nano-scale quantum dots with good optical properties can be fabricated using a natural growth mode of strained layer semiconductors.[21] When InAs is deposited on GaAs it initially grows as a strained two-dimensional sheet, but beyond some critical thickness, tiny islands like those shown in Fig.1a form in order to minimize the surface strain. Overgrowth of the islands leads to the coherent incorporation of $In_xGa_{1-x}As$ dots into the crystal structure of the device, as can be seen in the cross-sectional image of Fig.1c. The most intensively studied are small InAs dots on GaAs emitting around 900-950nm at low temperatures, which can be conveniently measured with low noise Si single photon detectors.

A less desirable feature of the self-organising technique is that the dots form at random positions on the growth surface. However, recently considerable progress has been made on controlling the dot position (Fig.1b) within the device structure by patterning nanometer sized pits on the growth surface.[22,23]

As InGaAs has a lower energy bandgap than GaAs, the quantum dot forms a potential trap for electrons and holes. If sufficiently small, the dot contains just a few quantised levels in the conduction and valence bands, each of which holds two electrons or holes of opposite spin. Illumination by a picosecond laser pulse excites electrons and holes which rapidly relax to the lowest lying energy states either side of the bandgap. A quantum dot can thus capture two electrons and two holes to form the biexciton state, which decays by a radiative cascade, as shown schematically in Fig.2a. One of the trapped electrons recombines with one of the holes and generates a first photon (called the biexciton photon, $X_2$). This leaves a single electron-hole pair in the dot (the exciton state), which subsequently also recombines to generate a second (exciton, $X$) photon. The biexciton and exciton photons have distinct energies, as can be seen in the low temperature photoluminescence spectrum of Fig.2a, due to the different Coulomb energies of their initial and final states. Often a number of other weaker lines can also be seen due to recombination of charged excitons which form intermittently when the dot captures an excess electron or hole.[24] Larger quantum dots, with several confined electron and hole levels, have a richer optical signature due to the large number of exciton complexes that can be confined.

High resolution spectroscopy reveals that the $X_2$ and $X$ transitions of a dot are in fact both doublets with linearly polarised components parallel to the [110] and [1-10] axes of the semiconductor crystal, labelled here H and V, respectively.[25,26] The origin of this polarisation is an asymmetry in the electron-hole exchange interaction of the dot which produces a splitting of the exciton spin states. The asymmetry derives from an elongation of the dot along one crystal axis and in-built strain in the crystal. It mixes the exciton eigenstates of a symmetric dot with total z-spin $J_z$ = +1 and -1 into symmetric and anti-symmetric combinations, which couple to two H or two V polarised photons, respectively, as shown in Fig.2.

The exciton state of the dot has a typical lifetime of ~1ns, which is due purely to radiative decay. As this is much longer than the duration of the exciting laser pulse, or the lifetime of the photo-excited carrier population in the surrounding semiconductor, only one $X$ photon can be emitted per laser pulse. This can be proven, as first reported[27] by Peter Michler, Atac Imamoglu and their colleagues in Santa Barbara, by measuring the second order correlation function, $g^{(2)}(\tau)$ of the exciton photoluminescence,[28,29] see text box. In fact each of the exciton complexes of the dot generates at most one photon per excitation cycle, which allows single-photon emission from also the biexciton or charged exciton transitions.[30]

Cross-correlation measurements[31,32,33] between the $X$ and $X_2$ photons confirm the time correlation expected for the cascade in Fig.2a, ie the $X$ photon follows the $X_2$ one. Indeed the shape of the cross-correlation function for both CW and pulsed excitation can be accurately described with a simple rate equation model and the experimentally measured $X$ and $X_2$ decay rates.[34]

## Semiconductor Microcavities

A major advantage of using self-assembled quantum dots for single-photon generation is that they can be easily incorporated into cavities using standard semiconductor growth and processing techniques. Cavity effects are useful for directing the emission from the dot into an experiment or application, as well as for modifying the photon emission dynamics.[35][36] Purcell[37] predicted enhanced spontaneous emission from a source in a cavity when its energy coincides with that of the cavity mode, due to the greater density of optical states to emit into. For an ideal cavity, in which the emitter is located at the maximum of the electric field with its dipole aligned with the local electric field, the enhancement in decay rate is given by $F_p = (3/4\pi^2) (\lambda/n)^3 Q/V$, where Q is the quality factor, a measure of the time a photon is trapped in the cavity, and V is the effective mode volume. Thus high photon collection efficiency, and simultaneously fast radiative decay, requires small cavities with highly reflecting mirrors and a high degree of structural perfection. However, without controlling the location of the dot in the cavity, as discussed below, it may be difficult to achieve the full enhancement predicted by the Purcell formula.

Figure 3 shows images of some of the single quantum dot cavity structures that have proven most successful. Pillar microcavities, formed by etching cylindrical pillars into semiconductor Bragg mirrors placed either side of the dot layer, have shown large Purcell enhancements and have a highly directional emission profile, thus making good single-photon sources.[38][39][40][41] Purcell factors of around 6 have been measured directly,[40][41] through the rate of cavity-enhanced radiative decay compared to that of a dot without cavity, implying a coupling to the cavity mode of $\beta=F_p/(1+F_p)>85\%$, if we assume the leaky modes are unaffected by the cavity. However, the experimentally determined photon collection efficiency, which is a more pertinent parameter for applications, is typically ~10%, due the fact that not all the cavity mode can be coupled into an experiment and scattering of the mode by the rough pillar edges. We can expect that the photon collection efficiency will increase with improvements to the processing technology or new designs of microcavity.

Another means of forming a cavity is to etch a series of holes in a suspended slab of semiconductor, so as to form a lateral variation in the refractive index which creates a forbidden energy gap for photonic modes in which light cannot propagate.[42] Photons can then be trapped in a central irregularity in this structure: usually an unetched portion of the slab. Such photonic bandgap defect cavities have been fabricated in Si with Q values approaching $10^6$.[43][44] High quality active cavities have also been demonstrated in GaAs containing InAs quantum dots.[45][46][47][48] A radiative lifetime of 86 ps, corresponding to a Purcell factor of $F_p\sim12$, has been reported.[47] Very recently a lifetime of 60ps was measured for a cavity in the strong coupling regeme.[48]

If the Q-value is sufficiently large, the system enters the strong coupling regime where the excitation oscillates coherently between an exciton in the dot and a photon in the cavity. The spectral signature of strong coupling, an anti-crossing between the dot line and the cavity mode, has been observed for quantum dots in pillar microcavities,[49] photonic bandgap defect cavities,[50] microdisks[51] and microspheres.[52] It has been demonstrated for atom cavities that strong coupling allows the deterministic generation of single-photons.[53][54] Single-photon sources in the strong coupling regime can be expected to have very high extraction efficiencies and be time-bandwidth limited.[55] Encouragingly single-photon emission has been reported recently for a dot in a strongly coupled pillar microcavity.[56]

Another interesting recent development is the ability to locate a single quantum dot within the cavity, as this ensures the largest possible coupling and removes background emission, as well as other undesirable effects, due to other dots in the cavity. Above we discussed techniques to control the dot position on the growth surface. The other way is to position the cavity around the dot. One technique combines micro-photoluminescence spectroscopy to locate the dot position, with in-situ laser photolithography to pattern markers on the wafer surface.[57] An alternative involves growing a vertical stack of dots so that their location can be revealed by scanning the wafer surface,[58] as shown in Fig.3. Recently this technique has allowed larger coupling energies for a single dot in a photonic bandgap defect cavity.[48]

**Photon Indistinguishability**

Cavity effects are important for rendering different photons from the source indistinguishable, which is essential for many applications in quantum information. When two identical photons are incident simultaneously on the opposite input ports of a 50/50 beamsplitter, they will always exit via the same output port,[59] as shown schematically in Fig.4a. This occurs because of a destructive interference in the probability amplitude of the final state in which one photon exits through each output port. The amplitude of the case where both photons are reflected exactly cancels with that where both are transmitted, due to the $\pi/2$ phase change upon reflection, provided the two photons are entirely identical.

Two-photon interference of two single-photons emitted successively from a quantum dot in a weakly-coupled pillar microcavity was first reported by the Stanford group.[60] Fig. 4b shows a schematic of their experiment. Notice the reduction of the co-incidence count rate measured between detectors in either output port, when the two photons are injected simultaneously (Fig.4c). The dip does not extend completely to zero, indicating that the two photons sometime exit the beamsplitter in opposite ports. The measured reduction in co-incidence rate at zero delay of 69%, implies an overlap for the single-photon wavepackets of 0.81, after correcting for the imperfect single-photon visibility of the interferometer. Two-photon interference dips of 66% and 75% have been reported by Bennett et al[61] and Vauroutsis et al.[62] Similar results have been obtained for a single dot in a photonic bandgap defect cavity.[63]

This two-photon interference visibility is limited by the finite coherence time of the photons emitted by the quantum dot,[64] which renders them distinguishable. The depth of the dip in Fig.4c depends upon the ratio of radiative decay time to the coherence time of the dot, ie $R=2\tau_{decay}/\tau_{coh}$. When unity, the coherence time is limited by radiative decay and the source will display perfect 2-photon interference. The most successful approach thus far has been to extend $\tau_{coh}$ by resonant optical excitation of the dot and reduce $\tau_{decay}$ using the Purcell effect in a pillar microcavity, to values $R\sim1.5$. In the future higher visibilities may be achieved with a larger Purcell enhancement, using a single dot cavity in the strong-coupling regime or with electrical gating described in the next section.

A source of indistinguishable single-photons was used by Fattal et al to generate entanglement between post-selected pairs.[65,66] This involves simply rotating the polarisation of one of the photons incident on the final beamsplitter in Fig.4a by 90°. By post-selecting the results where the two photons arrive at the beamsplitter at the same time and where there is one photon in each output arm (labelled 1 and 2), the measured pairs should correspond to the Bell state

$$\psi^- = 1/\sqrt{2} \left( |H_1 V_2\rangle - |V_1 H_2\rangle \right) \quad \text{Eq.1}$$

Note that only if the two photons are indistinguishable and thus the entanglement is only in the photon polarisation, are the two terms in Eq1 able to interfere. Analysis of the density matrix published by Fattal et al[65] reveals a fidelity of the post-selected pairs to the state in Eq.1 of 0.69, beyond the classical limit of 0.5. This source of entangled pairs has an importance difference to that based on the biexciton cascade described below. Post-selection implies that the photons are destroyed when this scheme succeeds. This is a problem for some quantum information applications such as LOQC, but could be usefully applied to quantum key distribution.[65]

## Single-Photon LEDs

An early proposal for an electrical single-photon source by Kim et al[67] was based upon etching a semiconductor heterostructure displaying Coulomb blockade. However, the light emission from this etched structure was too weak to allow the second-order correlation function to be studied. Recently encouraging progress has been made towards the realisation of a single-photon source based on quantising a lateral electrical injection current.[68,69] However the most successful approach so far has been to integrate self-assembled quantum dots into conventional p-i-n doped junctions.

In the first report of electrically-driven single-photon emission by Yuan et al,[70] the electroluminescence of a single dot was isolated by forming a micron-diameter emission aperture in the opaque top contact of the p-i-n diode. Fig.5a shows an improved emission aperture single-photon LED after Bennett et al,[71] which incorporates an optical cavity formed between a high reflectivity Bragg mirror and the semiconductor/air interface in the aperture. This structure forms a weak cavity, which enhances the measured collection efficiency 10-fold compared to devices without a cavity.[72]

Single-photon pulses are generated by exciting the diode with a train of short voltage pulses. The second order correlation function $g^{(2)}(\tau)$ of either the *X* or *$X_2$* electroluminescence (Fig.5c) shows the suppression of the zero delay peak indicative of single-photon emission.[71] The finite rate of multi-photon pulses is due mostly to background emission from layers other than the dot, which is also seen for non-resonant optical excitation. Electrical contacts also allow the temporal characteristics of the single-photon source to be tailored. By applying a negative bias to the diode between the electrical injection pulses, Bennett et al[73] reduced the jitter in the photon emission time <100ps. This allowed the repetition rate of the single-photon source to be increased to 1.07GHz (Fig.5d) while retaining good single-photon emission characteristics (Fig.5e). Electrical gating could provide a technique for producing time-bandwidth-limited single-photons from quantum dots.

Another promising approach is to aperture the current flowing through the device.[74,75] This is achieved by growing a thin AlAs layer within the intrinsic region of the p-i-n junction and later exposing the mesa to wet oxidation in a furnace, converting the AlAs layer around the outer edge of the mesa to insulating Aluminium oxide. By careful control of the oxidation time, a μm-diameter conducting aperture can be formed within the insulating ring of AlOx. Such structures have the advantage of exciting just a single dot within the structure, thereby reducing the amount of background emission. The oxide annulus also confines the optical mode laterally within the structure, potentially allowing high photon extraction efficiency.

Altering the nanostructure or materials that comprise the quantum dot allows considerable control over the emission wavelength and other characteristics. Most of the experimental work done so far has concentrated on small InAs quantum dots emitting around 900-950nm, as these have well understood optical properties and can be detected with low noise Si single-photon detectors. On the other hand the shallow confinement potentials of this system means they emit only at low temperatures. At shorter wavelengths optically-pumped single-photon emission has been demonstrated at ~350nm using GaN/AlGaN,[76] 500nm using CdSe/ZnSSe[77] and 682nm InP/GaInP[78] quantum dot. The former two systems have been shown to operate at 200K.

It is very important for quantum communications to develop sources at longer wavelengths in the fibre optic transmission bands at 1.3 and 1.55μm. This may be achieved using InAs/GaAs heterostructures by depositing more InAs to form larger quantum dots. These larger dots offer deeper confinement potentials than those at 900nm and thus often display room temperature emission.[79] Optically pumped single-photon emission at telecom wavelengths has been achieved using a number of techniques to prepare low densities of longer wavelength dots, including a bimodal growth mode in MBE to form low densities of large dots,[80] ultra-low growth rate MBE[81] and MOCVD.[82] Recently, the first electrically-driven single-photon source at a telecom wavelength has been demonstrated.[83]

## Generation of Entangled Photons

By collecting both the $X_2$ and $X$ photons emitted by the biexciton cascade, a single quantum dot may also be used as a source of photon pairs. Polarisation correlation measurements on these pairs discovered that the two photons were classically-correlated with the same linear polarisation.[84,85,86] This occurs because the cascade can proceed via one of two intermediate exciton spin states, as described above and shown in Fig.2a, one of which couples to two H- and the other two V-polarised photons. The emission is thus a statistical mixture of $|H_{X2}H_X\rangle$ and $|V_{X2}V_X\rangle$, although exciton spin scattering during the cascade (discussed below) ensures there are also some cross-polarised pairs.

The spin splitting[87,88] of the exciton state of the dot distinguishes the H and V polarised pairs and prevents the emission of entangled pairs predicted by Benson et al.[89] If this splitting could be removed, the H and V components would interfere in appropriately designed experiments. The emitted 2-photon state should then be written as a superposition of HH and VV, which can be recast in either the diagonal (spanned by D, A) or circular ($\sigma^+$, $\sigma^-$) polarisation bases, ie

$$\begin{aligned}\Phi^+ &= 1/\sqrt{2}\ (|H_{X2}\ H_X\rangle + |V_{X2}\ V_X\rangle) \\ &= 1/\sqrt{2}\ (|D_{X2}\ D_X\rangle + |A_{X2}\ A_X\rangle) \\ &= 1/\sqrt{2}\ (|\sigma^+_{X2}\ \sigma^-_X\rangle + |\sigma^+_{X2}\ \sigma^-_X\rangle)\end{aligned}\quad \text{Eq.2.}$$

Equal weighting of the HH and VV terms assumes the source to be unpolarised, as indicated by experimental measurements.

Eq.2 suggests that, for zero exciton spin splitting, the biexciton cascade generates entangled photon pairs, similar to those seen for atoms.[90] Entanglement of the $X$ or $X_2$ photons was recently observed experimentally for the first time by Stevenson, Young and co-workers,[91,92] using two different schemes to cancel the exciton spin splitting. An alternative approach by Akopian et al,[93] using dots with finite exciton splitting, post-selects photons emitted in a narrow spectral band where the two polarisation lines overlap.

The exciton spin splitting depends on the exciton emission energy, tending to zero for InAs dots emitting close to 1.4eV and then inverting for higher emission energy.[94,95] These correspond to shallow quantum dots for which the carrier wavefunctions extend into the barrier material reducing the electron-hole exchange. Zero splitting can be achieved by either careful control of the growth conditions to achieve dots emitting close to the desired energy, or by annealing samples emitting at lower energy.[94] The exciton spin splitting may be continuously tuned by applying a magnetic field in the plane of the dot.[96] It has been observed that the signatures of entanglement then appear only when the exciton splitting is close to zero.[91] Other promising schemes to tune the exciton splitting are now emerging, including application of strain[97] and electric field.[98,99]

Figure 6a plots polarisation correlations reported by Young et al[92] for a dot with zero exciton splitting (by control of the growth conditions). Pairs emitted in the same cascade (ie zero delay) shows a very striking positive correlation (co-polarisation) measuring in either, rectilinear or diagonal bases and anti-correlation (cross-polarisation) when measuring in circular basis. This is exactly the behaviour expected for the entangled state of Eq.2. In contrast, a dot with finite splitting shows polarisation correlation for the rectilinear basis only, with no correlation for diagonal or circular measurements, see Figure 6b. The strong correlations seen for all three bases in Fig.6a could not be produced by any classical light source or mixture of classical sources and is proof that the source generates entangled photons. The measured[92] two-photon density matrix (Fig.6c) projects onto the expected $1/\sqrt{2}\ (|H_{X2}\ H_X\rangle + |V_{X2}\ V_X\rangle)$ state with fidelity (ie probability) $0.702 \pm 0.022$, exceeding the classical limit (0.5) by 9 standard deviations.

Two processes contribute to the 'wrongly' correlated pairs which impair the fidelity of the entangled photon source. The first of these is due to background emission from layers in the sample other than the dot. This background emission, which is unpolarised and dilutes the entangled photons from the dot, limited the fidelity observed in the first report[91] of triggered entangled photon pairs from a quantum dot and has been subsequently reduced with better sample design.[92] The second mechanism, which is an intrinsic feature of the dot, is exciton spin scattering during the biexciton cascade. It is interesting that this process does not seem to depend strongly upon the exciton spin splitting. It may be reduced by suppressing the scattering using resonant excitation or alternatively using cavity effects to reduce the time required for the radiative cascade.

## Outlook

The past several years have seen remarkable progress in quantum light generation using semiconductor devices. However, despite considerable progress many challenges still remain. The structural integrity of cavities must continue to improve, thereby enhancing quality factors. This, combined with the ability to reliably position single dots within the cavity, will further enhance photon collection efficiencies and the Rabi energy in the strong coupling regime. It is also important to realise all the benefits of these cavity effects in more practical electrically-driven sources. Meanwhile bandstructure engineering of the quantum dots will allow a wider range of wavelengths to be accessed for both single and entangled photon sources, as well as structures that can operate at higher temperatures. Techniques for fine tuning the characteristics of individual emitters will also be important.

One of the most interesting aspects of semiconductor quantum optics is that we may be able to use quantum dots not only as quantum light emitters, but also as the logic and memory elements which are required in quantum information processing. Although LOQC is scalable theoretically, quantum computing with photons would be much easier with a useful single-photon non-linearity. Such non-linearity may be achieved with a quantum dot in a cavity in the strong coupling regime. Encouragingly strong coupling of a single quantum dot with various type of cavity has already been observed in the spectral domain. Eventually it may even be possible to integrate photon emission, logic, memory and detection elements into single semiconductor chips to form a photonic integrated circuit for quantum information processing.

The author would like to thank Mark Stevenson, Robert Young, Anthony Bennett, Martin Ward and Andy Hudson for their useful comments during the preparation of the manuscript and the UK DTI "Optical Systems for Digital Age", EPSRC and EC Future and Emerging Technologies programmes for supporting research on quantum light sources.

# TextBox : Photon Correlation Measurements

The photon statistics of light can be studied via the second order correlation function, $g^{(2)}(\tau)$, which describes the correlation between the intensity of the light field with that after a delay $\tau$ and is given by[100]

$$g^{(2)}(\tau) = \frac{<I(t)I(t+\tau)>}{<I(t)>^2}$$

This function can be measured directly using the Hanbury-Brown and Twiss[101] interferometer, comprising a 50/50 beamsplitter and two single-photon detectors, shown in the figure. For delays much less than the average time between detection events (ie for low intensities), the distribution in the delays between clicks in each of the two detectors is proportional to $g^{(2)}(\tau)$.

For a continuous light source with random emission times, such as an ideal laser or LED, $g^{(2)}(\tau)=1$. It shows there is no correlation in the emission time of any two photons from the source. A source for which $g^{(2)}(\tau=0)>1$ is described as 'bunched' since there is an enhanced probability of two photons being emitted within a short time interval. Photons emitted by quantum light sources are typically 'anti-bunched', ($g^{(2)}(\tau=0)<1$) and tend to be separated in time.

In communication and computing systems, we are interested in pulsed light sources, for which the emission occurs at times defined by an external clock. In this case $g^{(2)}(\tau)$ consists of a series of peaks separated by a clock period. For an ideal single-photon source, the peak at zero time delay is absent, $g^{(2)}(\tau=0)=0$; as the source cannot produce more than one photon per excitation period, clearly the two detectors cannot fire simultaneously.

The figure shows $g^{(2)}(\tau)$ recorded for resonant pulsed optical excitation of the *X* emission of a single quantum dot in a pillar microcavity. Notice the almost complete absence of the peak at zero delay: the definitive signature of a single-photon source. The weak peak seen at $\tau=0$ demonstrates that the rate of two-photon emission is 50 times less than that of an ideal laser with the same average intensity. The bunching behaviour observed for the finite delay peaks is explained by intermittent trapping of a charge carrier in the dot.[102] This trace was taken for quasi-resonant laser excitation of the dot which avoids creating carriers in the surrounding semiconductor. For higher energy laser excitation, the suppression in $g^{(2)}(0)$ is typically reduced indicating occasional 2-photon pulses due to emission from the layers surrounding the dot, but can be minimised with careful sample design.

**Figure textbox**: (a) Schematic of the set-up used for photon correlation measurements, (b) second order correlation function of the exciton emission of a single dot in a pillar microcavity.

# Figure Captions

**Figure 1**: Self assembled quantum dots (a) Image of a layer of InAs/GaAs self assembled quantum dots recorded in an Atomic Force Microscope (AFM). Each yellow blob corresponds to a dot with typical lateral diameters of 20-30nm and a height of 4-8nm. (b) AFM image[23] of a layer of InAs quantum dots whose locations have been seeded by a matrix of nanometer sized pits patterned onto the wafer surface. Under optimal conditions up to 60% of the etch pits contain a single dot (Courtesy of P Atkinson & D A Ritchie, Cambridge). (c) Cross-sectional STM image of an InAs dot inside a GaAs device (Courtesy of P. Koenraad, Eindhoven).

**Figure 2**: Optical spectrum of a quantum dot. (a) Schematic of the biexciton cascade of a quantum dot. (b) Typical photoluminescence spectrum of a single quantum dot showing sharp line emission due to the biexciton $X_2$ and exciton $X$ photon emitted by the cascade. The inset shows the polarisation splitting of the transitions originating from the spin splitting of the exciton level.

**Figure 3**: SEM images of semiconductor cavities, including pillar microcavities (a)[56] and (b), microdisk (c)[51] and photonic bandgap defect cavities (d)[47], (e) and (f).[48] (Structures fabricated at Univ Wuerzburg (a), CNRS-LPN (UPR-20), Marcoussis (b, c, e), Univ Cambridge (d), UCSB/ETHZ Zurich (f))

**Figure 4**: Two Photon Interference. (a) If the two photons are indistinguishable, the two outcomes resulting in one photon in either arm interfere destructively. This results in the two photons always exiting the beamsplitter together. (b) Schematic of an experiment using two photons emitted successively from a quantum dot, (c) experimental data showing suppression of the co-incidence rate in (b) when the delay between input photons is zero due to two-photon interference.[60] (Courtesy of Y Yamamoto, Stanford Univ.)

**Figure 5**: Electrically driven single-photon emission. (a) Schematic of a single-photon LED. (b) Electroluminescence spectra of the device. Notice the spectra are dominated by the exciton $X$ and biexciton $X_2$ lines, which have linear and quadratic dependence on drive current, respectively. Other weak lines are due to charged excitons. (c) second order correlation function recorded for the exciton (i) and biexciton (ii) emission lines, (d) time-resolved electroluminescence from a device operate with a 1.07GHz repetition rate, (e) measured (i) and modelled (ii) second order correlation function of the biexciton electroluminescence at 1.07GHz. (adapted from Refs. 71 and 73)

**Figure 6**: Generation of entangled photons by a quantum dot. (a) Degree of correlation measured for a dot with exciton polarisation splitting S=0 μeV in linear (i), diagonal (ii) and circular (iii) polarisation bases as a function of the delay between the $X$ and $X_2$ photons (in units of the repetition cycle). The correlation is defined as the rate of co-polarised pairs minus the rate of cross-polarised pairs divided by the total rate. Notice that the values at finite delay show no correlation, as expected for pairs emitted in different laser excitation cycles. More interesting are the peaks close to zero time delay, corresponding to $X$ and $X_2$ photon emitted from the same cascade. The presence of strong correlations for all three types of measurement for the dot with zero exciton splitting can only be explained if the $X$ and $X_2$ polarisations are entangled. (b) Degree of correlations measured for the dot in (a) subject to in-plane magnetic field so as to produce an exciton polarisation splitting of S=25 μeV. Notice that the correlation in diagonal and circular bases have vanished, indicating only classical correlations at finite splitting. (c) Two-photon density matrix of the device emission in (a). The strong off-diagonal terms appear due to entanglement. (adapted from Ref 92)

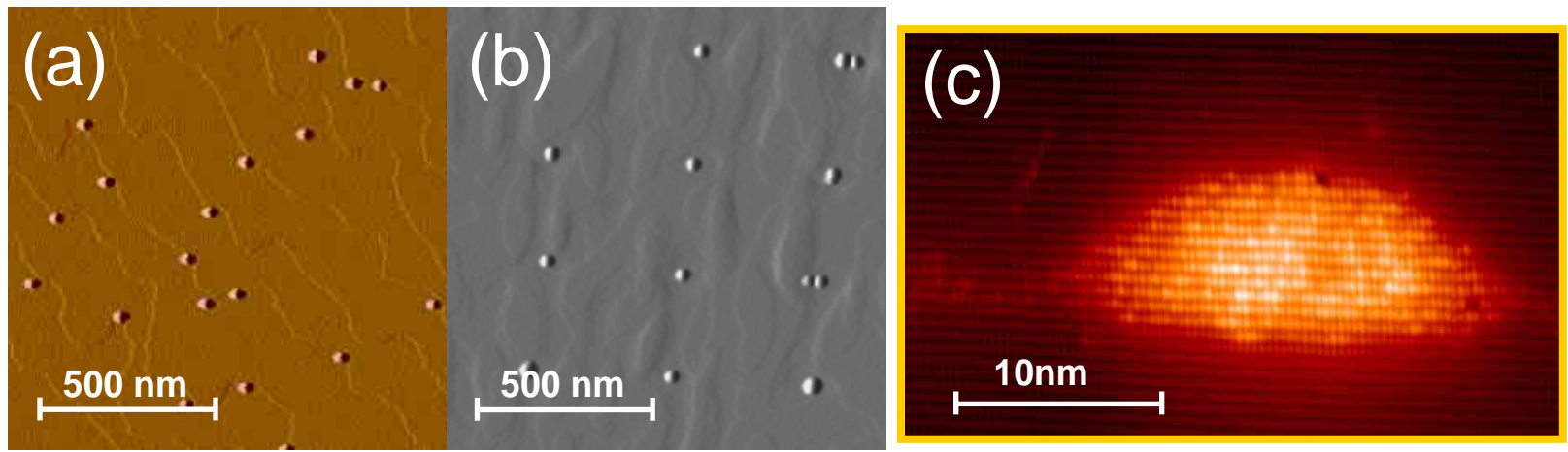

Fig. 1

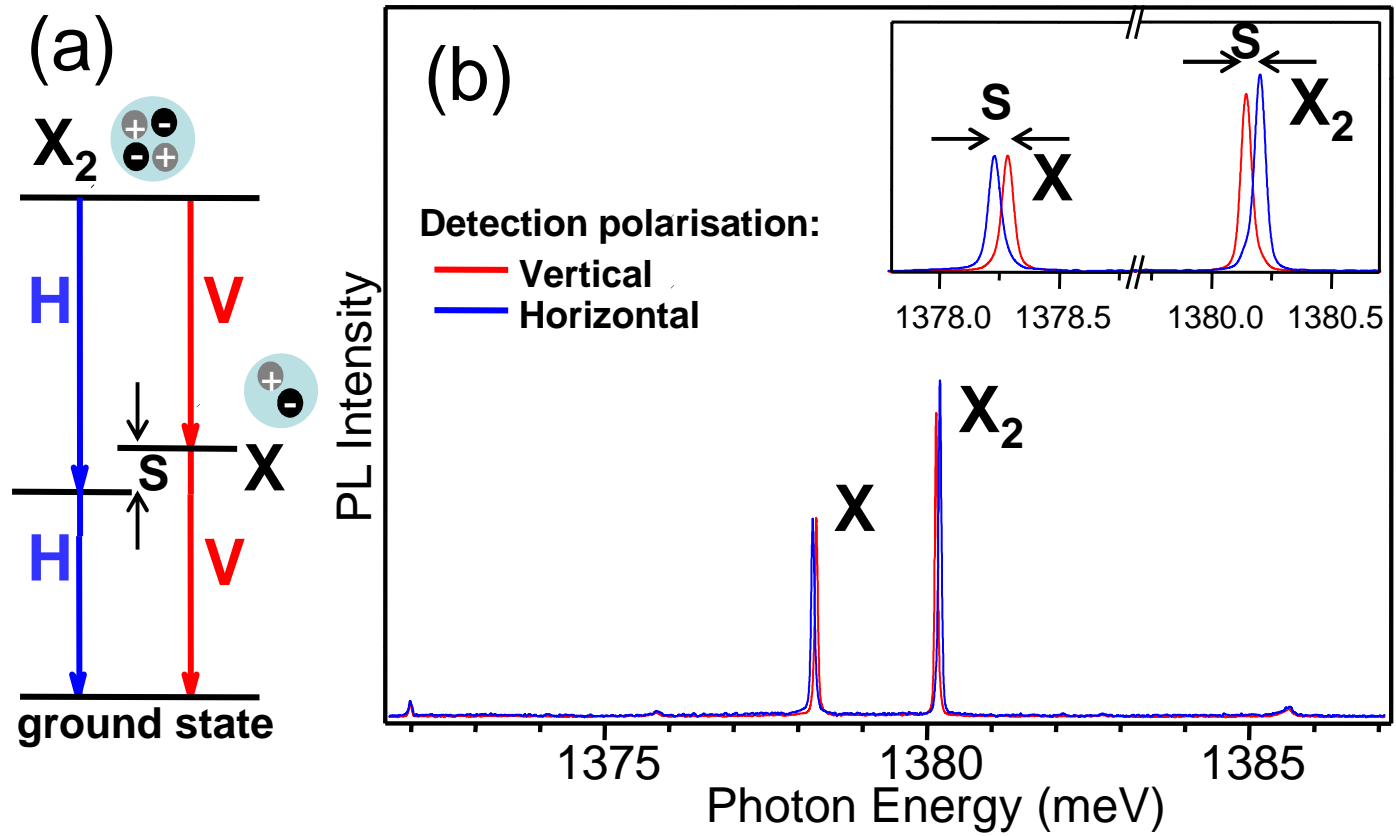

**Fig. 2**

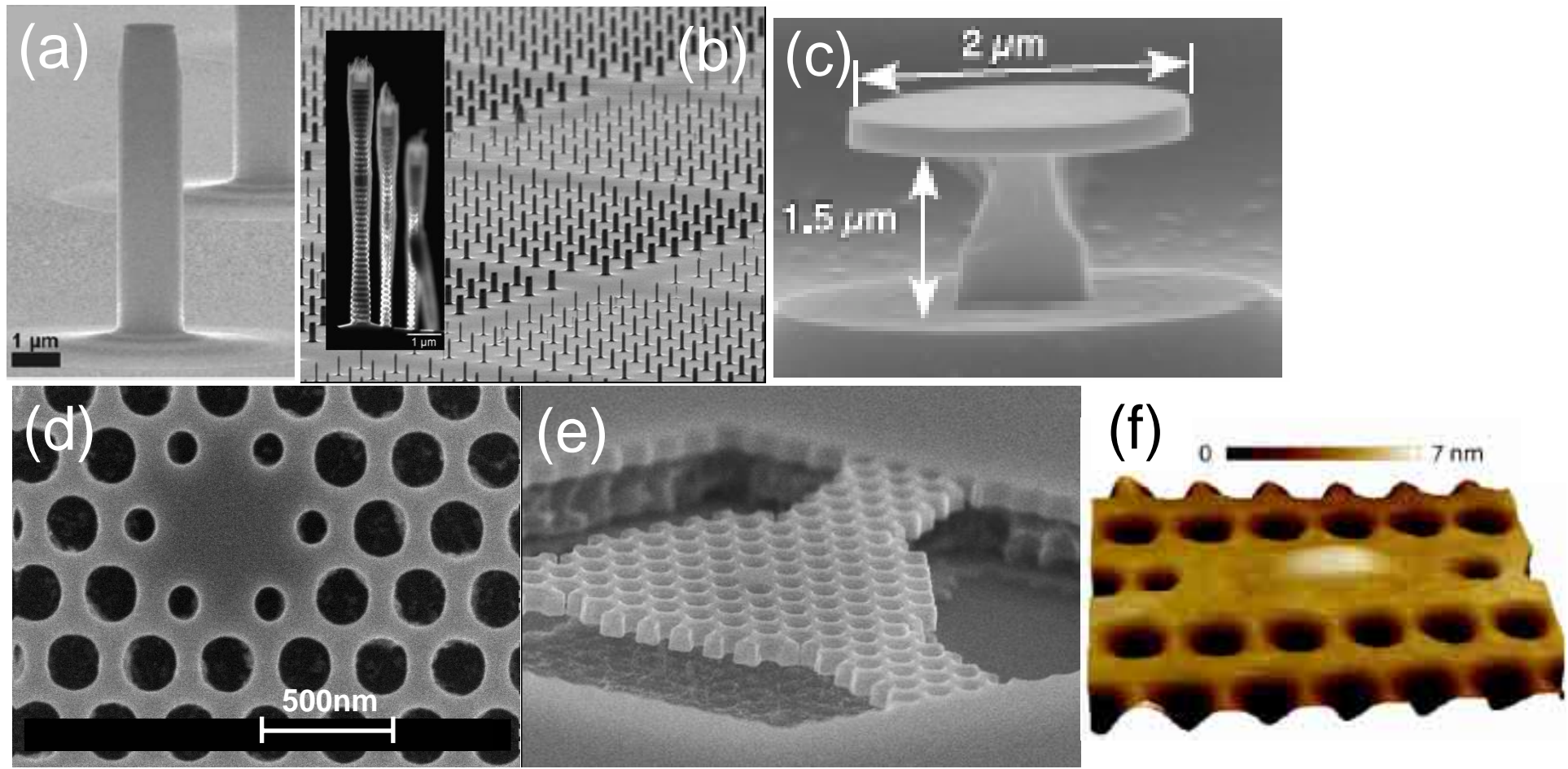

**Fig. 3**

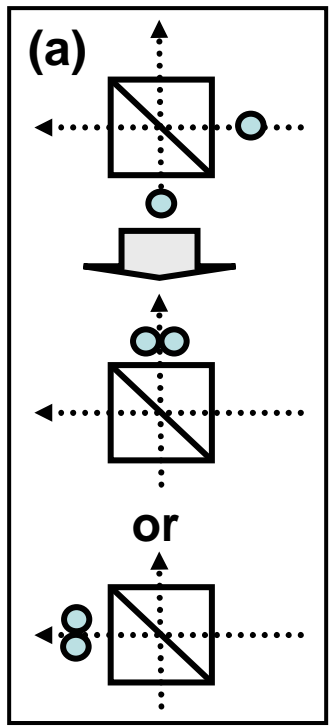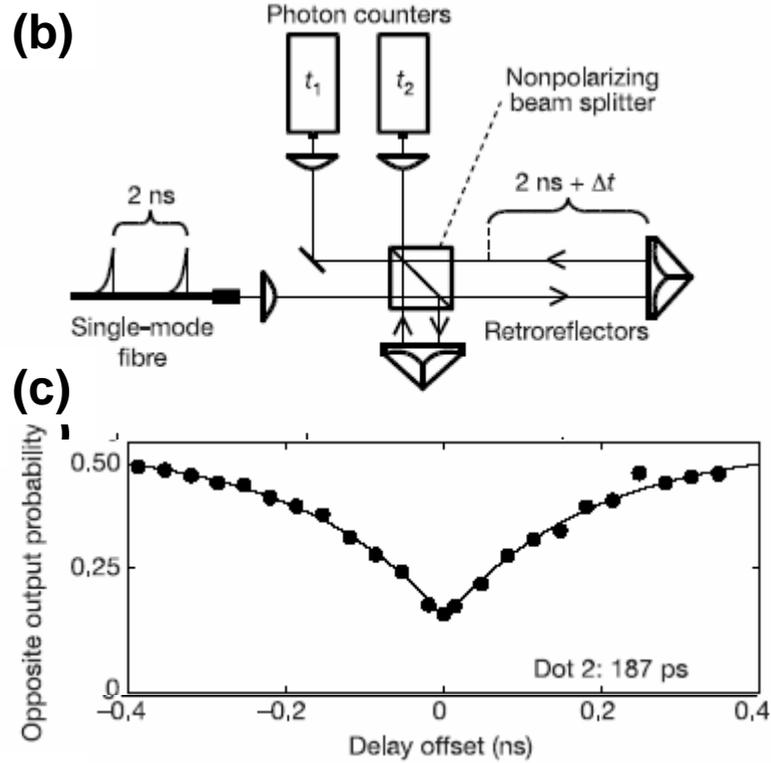

Fig. 4

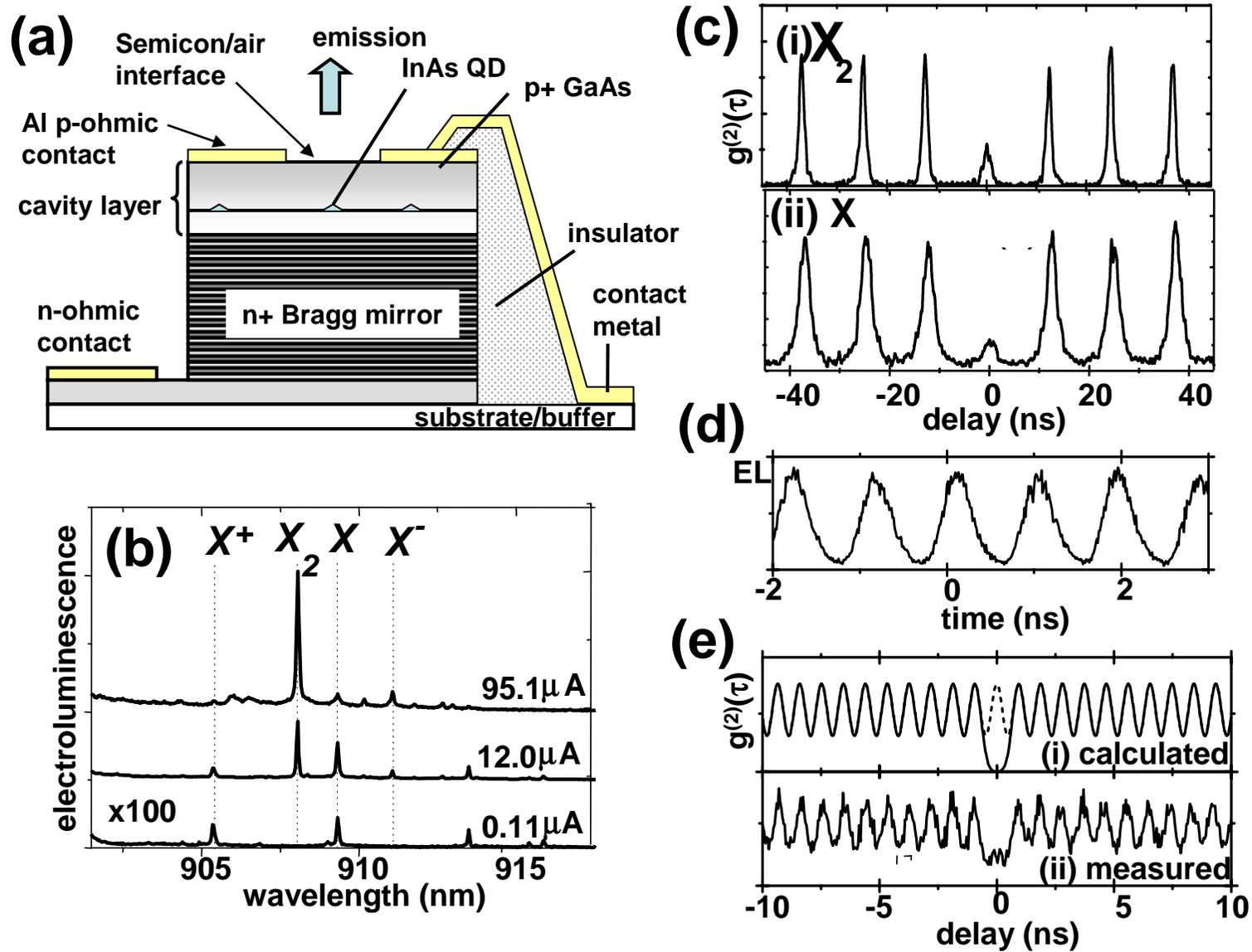

Fig. 5

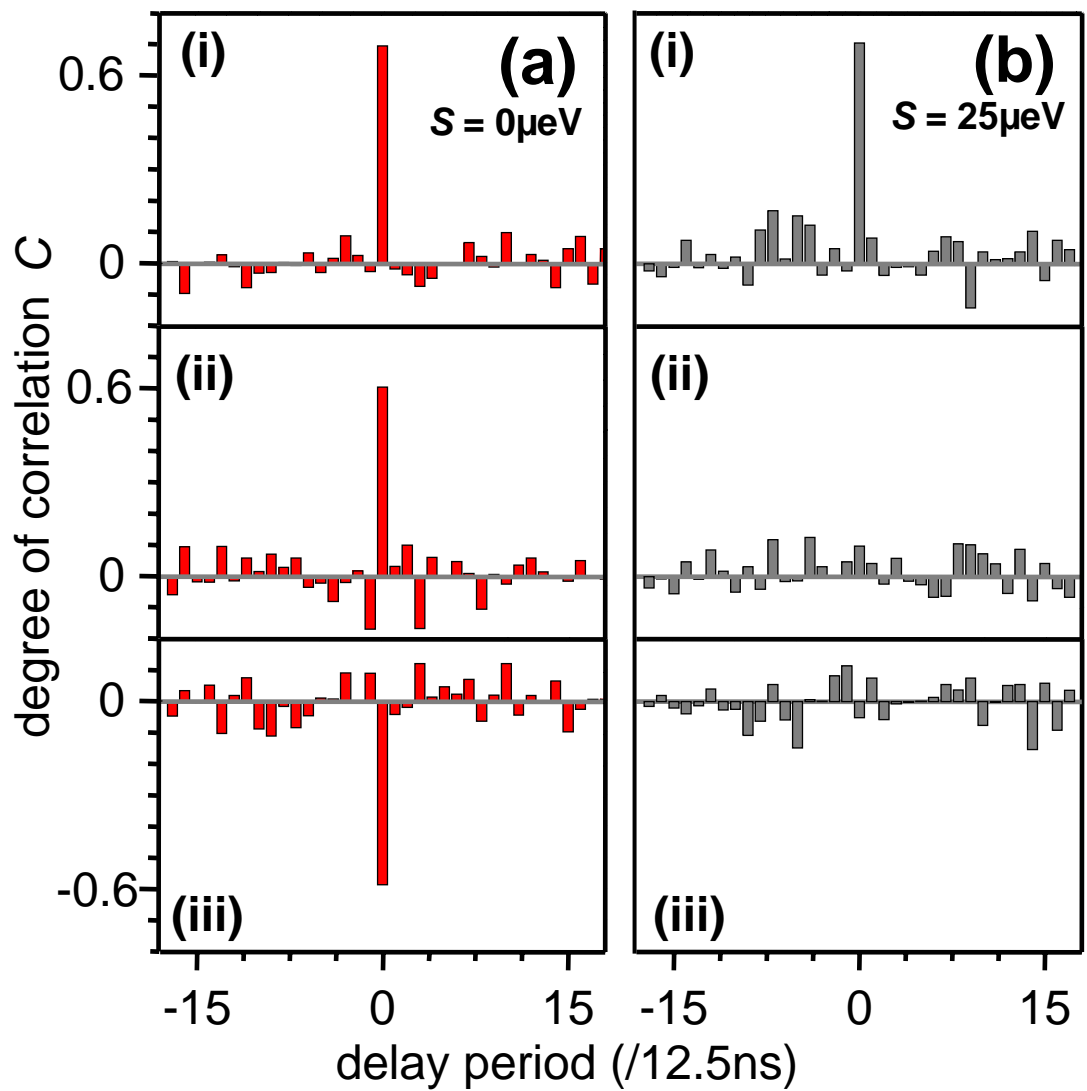
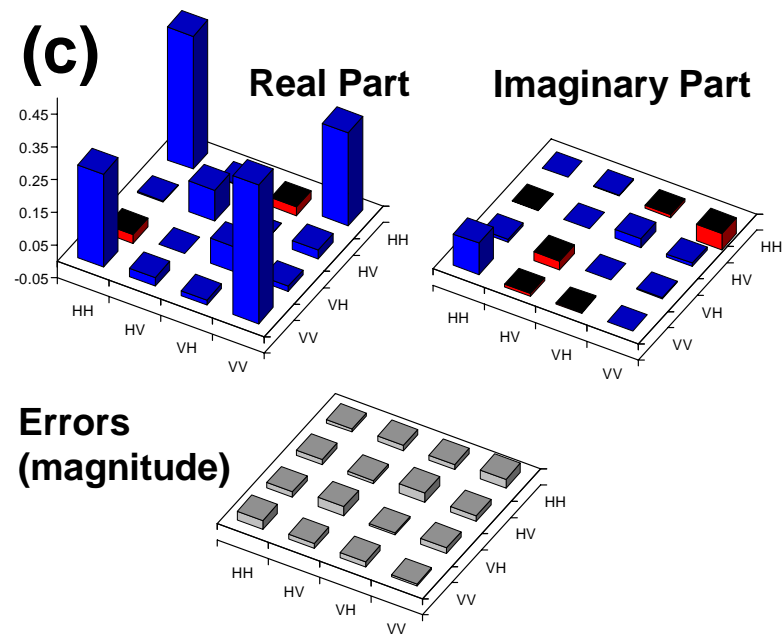

Fig. 6

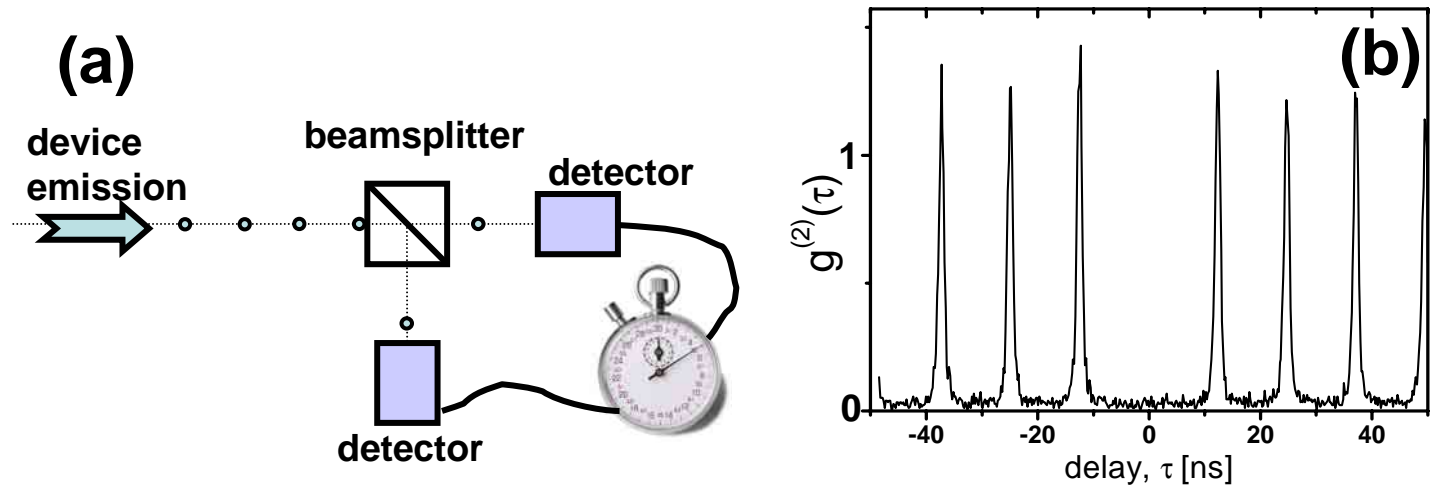

**Fig. textbox**